\documentclass[
reprint,
superscriptaddress,
frontmatterverbose, 
showpacs,
preprintnumbers,
nofootinbib,
amsmath,
amssymb,
prl,
onecolumn,
floatfix
]{revtex4-2}

\usepackage[inline]{enumitem}
\usepackage[utf8]{inputenc}
\usepackage[normalem]{ulem}
\usepackage{graphicx}
\usepackage{dcolumn}
\usepackage{bm}
\usepackage{color}
\usepackage[dvipsnames]{xcolor}
\usepackage[
    colorlinks = true,
    linkcolor = BlueViolet,
    anchorcolor = purple,
    citecolor = purple,
    filecolor = purple,
    urlcolor = BlueViolet]{hyperref}
    
\usepackage{url}
\usepackage{xspace}
\usepackage{slashed}
\usepackage{multirow,bigstrut}
\usepackage{mathrsfs} 
\usepackage{relsize,amsmath}
\usepackage[geometry]{ifsym}
\usepackage{amssymb}
\usepackage{pifont}
\usepackage{physics}

\usepackage{cleveref}

\usepackage{rotating}
\usepackage{ragged2e}

\newcommand{\nano}{NANOGrav }

\newcommand{\titleprl}[1]{\vspace{0.3cm} \noindent \textit{\textbf{#1}}:}

\usepackage{siunitx}

\begin{document}

\title{Boiling away asymmetries: low-scale phase transitions, gravitational waves and leptogenesis}

\author{Leonardo Grimaldi}
\email{leonardo.grimaldi3@studio.unibo.it}
\affiliation{Dipartimento di Fisica e Astronomia, Universit\`a di Bologna, via Irnerio 46, 40126 Bologna, Italy}
\affiliation{Max Planck Institute for Gravitational Physics (Albert Einstein Institute), D-14476 Potsdam, Germany}

\author{Michele Lucente}
\email{michele.lucente@unibo.it}
\affiliation{Dipartimento di Fisica e Astronomia, Universit\`a di Bologna, via Irnerio 46, 40126 Bologna, Italy}
\affiliation{INFN, Sezione di Bologna, viale Berti Pichat 6/2, 40127 Bologna, Italy}

\author{Silvia Pascoli}
\email{silvia.pascoli@unibo.it}
\affiliation{Dipartimento di Fisica e Astronomia, Universit\`a di Bologna, via Irnerio 46, 40126 Bologna, Italy}
\affiliation{INFN, Sezione di Bologna, viale Berti Pichat 6/2, 40127 Bologna, Italy}

\date{\today}

\begin{abstract}
Leptogenesis is one of the most popular mechanisms to account for the observed baryon asymmetry of the Universe. A generic feature of leptogenesis is a large separation of scales between the epoch of baryon asymmetry production (sphaleron freeze-out at temperature $T \sim 130$ GeV) and the one where it affects the big bang nucleosynthesis processes (BBN at $T \sim 1$ MeV). Any entropy release between these two epochs would lead to a dilution of previously produced relics, such as the baryons. Motivated by the recent evidence of a stochastic gravitational waves background (SGWB) in the nHz frequency range, we consider the case of supercooled first-order phase transition and we study the impact of the induced entropy dilution on the leptogenesis parameter space. We employ the Type-I seesaw with 2 right-handed neutrinos as benchmark scenario, and demonstrate that the viable leptogenesis parameter space is significantly reduced. Interestingly, the values of dilution predicted by the SGWB best fit points in several first order phase transition scenarios would completely exclude the leptogenesis parameter space testable by future experiments, thus establishing a phenomenological interconnection between leptogenesis, SGWB, first order phase transition and neutrino mass generation. Our analysis can be generalised to different leptogenesis models and entropy dilution mechanisms.
\end{abstract}

\maketitle

\titleprl{Introduction}
The origins of neutrino masses and the observed baryon asymmetry of the Universe (BAU) are outstanding puzzles in modern physics. Remarkably, both problems admit a viable explanation by extending the Standard Model (SM) via the addition of right-handed neutrino (RHN) fields: on the one hand, RHNs can give rise to small neutrino masses and sizeable lepton mixing via the seesaw mechanism~\cite{Minkowski:1977sc,Gell-Mann:1979vob,Yanagida:1979as,Mohapatra:1979ia}; on the other, their CP-violating interactions can source a BAU via the leptogenesis mechanism~\cite{Fukugita:1986hr,Covi:1996wh,Akhmedov:1998qx,Pilaftsis:2003gt,Giudice:2003jh,Asaka:2005pn,Abada:2006fw,Nardi:2006fx,Garbrecht:2011aw,Garny:2011hg,Frossard:2012pc,Canetti:2012kh,BhupalDev:2014oar,Kartavtsev:2015vto,Hernandez:2016kel,Abada:2018oly,Klaric:2020phc,Drewes:2021nqr}. 

The scale of the new physics, in particular the value of the RHN masses, is unconstrained from a theoretical point of view: viable leptogenesis realisations in scenarios with hierarchical RHN masses are possible from GUT~\cite{Buchmuller:2004nz} down to $10^6$ GeV~\cite{Davidson:2002qv,Buchmuller:2002rq,Buchmuller:2004nz,Racker:2012vw,Moffat:2018wke,Granelli:2021fyc,Granelli:2025cho}, while for degenerate RHN mass spectra resonant leptogenesis can account for the observed BAU and neutrino parameters with new physics at the GeV-TeV scale~\cite{Pilaftsis:2003gt,Pilaftsis:2005rv,Granelli:2020ysj}. If the lepton asymmetry is produced from the oscillation and scattering of the RHNs~\cite{Akhmedov:1998qx,Asaka:2005pn,Drewes:2012ma} (rather than from their decays) the preferred region is at the MeV-GeV scale~\cite{Asaka:2005pn,Shaposhnikov:2008pf,Asaka:2011wq,Drewes:2012ma,Canetti:2012kh,Shuve:2014zua,Garbrecht:2014bfa,Abada:2015rta,Hernandez:2015wna,Drewes:2016gmt,Hernandez:2016kel,Hambye:2016sby,Abada:2017ieq,Abada:2018oly,Domcke:2020ety}. Notice that, in general, a lepton asymmetry production can take place during both the freeze-in and freeze-out phases of RHN thermal history, with the relative importance of the processes determined by the specific model parameters~\cite{Klaric:2020phc,Klaric:2021cpi,Drewes:2021nqr}. In all cases, RHN are responsible for the production of a lepton asymmetry which is then converted into the observed baryon one by sphaleron effects~\cite{Linde:1977mm,Dimopoulos:1978kv,Kuzmin:1985mm,Rubakov:1996vz}. These latter processes emerge in the SM at the non-perturbative level and become ineffective below the temperature scale $T_* \sim 130$ GeV~\cite{DOnofrio:2014rug} (sphaleron decoupling). This implies that leptogenesis, as the source of the observed BAU, must have occurred at temperatures above 130~GeV and that any lepton asymmetry produced below this temperature will not affect the observed BAU value.

Because of the above considerations, any entropy release happening after the sphaleron decoupling would dilute the produced baryon asymmetry, changing the predictions of leptogenesis models. As a key example, we consider the expansion and subsequent reheating of the Universe due to a period of supercooled first order phase transition (FOPT)~\cite{Megevand:2007sv} below the electroweak scale. This can generically happen in dark sector models, in which new physics is included below the electroweak scale and in particular a dark scalar can break a dark symmetry, inducing a phase transition. For suitable values of the parameters, in particular the couplings of the new theory, such phase transition can be of first order, proceeding via the nucleation of bubbles of true vacuum and resulting in the production of a stochastic gravitational waves background (SGWB). This type of scenario has recently received significant attention, given that in 2023 several Pulsar Timing Array (PTA) collaborations reported the observation of an Hellings-Downs correlation in the residuals of the time of arrival of the tracked millisecond pulsars~\cite{NANOGrav:2023gor,EPTA:2023fyk,Reardon:2023gzh,Xu:2023wog}, providing evidence for the existence of a SGWB in the Universe in the nanoHertz frequency range. Although such a signal is expected from the merging of supermassive black holes binaries (SMBHB), the observed amplitude and spectral shape of the signal are in tension with astrophysical models~\cite{NANOGrav:2023hfp}, suggesting that a new cosmological source of SGWB at the nanoHertz frequency might be needed. Recent Bayesian analyses estimate $2.7\sigma$ preference for the FOPT explanation over SMBHB~\cite{Wang:2025wht}, even though it has  been shown that an improved noise model can mitigate tensions with astrophysical theoretical expectations for SMBHB~\cite{Goncharov:2024htb}.
In order to explain the observed signal, the FOPT needs to proceed very slowly, i.e. it has to be supercooled~\cite{Athron:2022mmm,Athron:2023mer,Ghosh:2023aum,Gouttenoire:2023bqy,Salvio:2023ynn,Salvio:2023blb,Conaci:2024tlc,Winkler:2024olr,Goncalves:2025uwh,Li:2025nja,Costa:2025csj,Balan:2025uke,Costa:2025pew}, and this leads to a short period of accelerated expansion and subsequent reheating that  dilutes any cosmological relic, including a BAU, already produced. This dilution has been discussed in the context of dark matter~\cite{Wainwright:2009mq,Davoudiasl:2015vba,DEramo:2017gpl,Hambye:2018qjv}.
 
In this letter, we show that leptogenesis models are strongly sensitive to possible entropy injections that may have occurred in the early Universe over a wide range of temperatures, $T \in [1 \textrm{ MeV}, 130 \textrm{ GeV}]$; in particular, testable leptogenesis mechanisms can be employed as trackers of the entropy evolution in the early Universe over the discussed temperature range, by comparing the BAU value with the asymmetry prediction resulting from the model parameters measured in laboratory experiments. We use as benchmark scenarios the SM extended with 2 RHN at the GeV mass scale for leptogenesis, and the entropy injection that results from cosmological FOPT at $\mathcal{O}(10)$ MeV temperatures (which is the typical scale needed to explain PTA observations), but we stress that our findings are generalisable to any testable leptogenesis and beyond the Standard Model (BSM) entropy injection mechanisms.

\titleprl{The baryon asymmetry and leptogenesis} 
The current BAU value is inferred by observations of the cosmic microwave background (CMB) and abundance of primordial elements produced during the big-bang nucleosynthesis (BBN) process, implying a ratio of baryons to photons density  $\eta \equiv n_b / n_\gamma = (6.143 \pm 0.190) \times 10^{-10}$ at the beginning of the BBN process~\cite{ParticleDataGroup:2024cfk}. 

Leptogenesis is a viable mechanism to explain such value. As we are interested in models in which it is possible to have some control over the predictions of the value of the asymmetry, we focus on ARS leptogenesis~\cite{Akhmedov:1998qx,Asaka:2005pn,Drewes:2012ma}. We consider the minimal SM extended with 2 RHN $N_I$, with $I = 1,2$:
\begin{equation}
    \mathcal{L}=\bar N_{I}i\slashed{\partial}N_{I}-\bar \ell_\alpha Y_{\alpha I} N_{I}\tilde{\Phi}-\frac{1}{2}\bar N_{I}^c {M_{R}}_{IJ} N_{J}+\mathrm{h.c.},
    \label{eq:lagrangian}
\end{equation}
where $Y_{\alpha I}$ is the Yukawa matrix, $\ell_\alpha$ are the leptonic doublets, with $\alpha= e,\mu, \tau$, $\tilde\Phi=i\sigma_2\Phi^*$ is the Higgs doublet, and $M_{R}$ is the Majorana mass matrix for the $N$ fields. In the absence of specific textures in the $Y$ and $M_R$ matrices, the Lagrangian in Eq.~(\ref{eq:lagrangian}) predicts an active-sterile mixing between SM neutrinos and new fields $N_I$ of order $\mathcal{O}(\sqrt{m_\nu / M_R})$, which is unobservable in current and planned experiments. This is not the case when the model features an approximate lepton-number conserving symmetry~\cite{Branco:1988ex,Ingelman:1993ve,Gluza:2002vs,Barr:2003nn,Shaposhnikov:2006nn,Kersten:2007vk,Ibarra:2010xw,Moffat:2017feq}, in which case the mixing can be sizeable while neutrino masses are protected from large radiative corrections; we thus work in such a scenario, employing the minimal flavour violation parametrisation in the Type-I seesaw with 2 RHN~\cite{Gavela:2009cd}. We note that in this case the two heavy neutrino mass states are nearly degenerate with mass $M$ and mass splitting $\Delta M$.

The baryon asymmetry generation in this model has been comprehensively analysed in~\cite{Asaka:2005pn,Shaposhnikov:2008pf,Canetti:2010aw,Canetti:2012kh,Abada:2015rta,Hernandez:2015wna,Hernandez:2016kel,Drewes:2016jae,Antusch:2017pkq,Eijima:2018qke,Klaric:2020phc,Granelli:2020ysj,Klaric:2021cpi,Hernandez:2022ivz,Sandner:2023tcg}. By imposing the constraints from neutrino oscillation experiments\footnote{In this model the lightest active neutrino is massless, thus the absolute neutrino mass scale is fixed by the oscillation frequencies.} there are 6 free parameters:  the heavy neutrino mass scale $M$, the heavy neutrino mass splitting $\Delta M$, the scale of Yukawa couplings $y$, the Dirac CP violating phase $\delta$, the Majorana CP violating phase $\phi$ and an additional phase associated to the heavy neutrino sector, $\theta$~\cite{Gavela:2009cd,Hernandez:2022ivz}. Experiments, on the other hand, can measure the individual heavy-neutrino masses and their mixings $|U_{\alpha}|^2\equiv\sum_{I=1,2}|\Theta_{\alpha I}|^2$ with the SM sector, where $\Theta^*=YvM_R^{-1}W^*$ is the active-sterile mixing matrix, and $W$ is the unitary matrix which diagonalizes $M_R$.

In the model in Eq.~(\ref{eq:lagrangian}) with RHN masses at the GeV scale, the lepton asymmetry is produced before the thermalisation of the heavy neutrino states: we assume initial vanishing abundance after reheating, such that the population of RHN is produced from inverse decays and $2 \rightarrow 2$ scatterings mediated by the small Yukawa couplings. Both the Yukawa couplings and the Majorana mass matrix can feature complex CP-violating phases. The RHN are produced as interaction eigenstates, generally misaligned with respect to the mass basis. This entails the coherent oscillation of RHN states. Lepton CP asymmetries result from the interference between the CP-violating coupling phases and the CP-conserving oscillation ones~\cite{Akhmedov:1998qx}. The generation of the baryon asymmetry is described by a system of quantum kinetic equations, which we implement following~\cite{Hernandez:2022ivz}. We employ the public code~\texttt{AMIQS}~\cite{Hernandez:2022ivz,Sandner:2023tcg} to numerically solve them and compute the baryon asymmetry at the sphaleron freeze-out temperature, $T_* = 131.7$ GeV~\cite{DOnofrio:2014rug}, performing a random scan of the parameter space sampling the following range of input values: $M \in [0.1, 100]$ GeV, $\Delta M/M \in [10^{-15}, 0.1]$, $y \in [10^{-8}, 10^{-4}]$ (all with a log-uniform distribution) and $\theta, \delta, \phi \in [0, 2\pi]$ for the CP-phases (with a linear uniform distribution). 

\titleprl{Dilution from FOPT} 
One of the possible mechanisms at the origin of the observed \nano signal~\cite{NANOGrav:2023gor} is a supercooled FOPT taking place in a dark sector~\cite{Bringmann:2023opz,Addazi:2023jvg,Ghosh:2023aum,Gouttenoire:2023bqy,Wang:2023bbc,Chen:2023bms,Conaci:2024tlc,Winkler:2024olr,Croon:2024mde,Goncalves:2025uwh,Costa:2025csj,Balan:2025uke}. Such a process is generally associated with an entropy release, which is expected to be sizeable to explain the \nano signal, thus significantly diluting any existing relic density. There are two mechanisms that release entropy during a FOPT: the first one is the change of the vacuum state of the Universe from a larger (false vacuum) to a lower (true vacuum) entropy state, with the resulting difference that is released to the thermal bath, resulting in a dilution $d_\textrm{vac}$ of the existing relic quantities. We adopt the minimal dark sector model from~\cite{Costa:2025csj} as a benchmark to estimate the importance of this contribution, and use the public software~\texttt{ELENA}~\cite{Costa:2025pew} to track the evolution of the entropy density $s$ of the vacuum states with the transition evolution. Here, the entropy $s$ is given by $s = P_f s_f + (1-P_f) s_t$, where $s_f$ ($s_t$) is the entropy density of the false (true) vacuum and $P_f$ is the fraction of the Universe in the false vacuum (cf. Supplemental Material for technical details). We find that such a transition results in moderate entropy releases, amounting at most to a $30\%$ increase between the start and the end of the transition for the considered benchmark points. The second entropy injection mechanism is the decay of the scalar field that drove the transition at the end of the process, that results into a reheating of the thermal bath up to the temperature $T_\textrm{reh} = \left(1 + \alpha\right)^\frac{1}{4}\ T_p$~\cite{Ellis:2018mja}, where $T_p$ is the percolation temperature of the transition and $\alpha$ is the transition strength. This reheating implies a dilution of the relic densities as $d_{\text{reh}}=(T_{\text{reh}}/ T_p )^3 = (1+\alpha)^{\frac34}$. For the minimal dark sector model considered here, four benchmark points are presented in~\cite{Costa:2025csj} and four more are added in ~\cite{Costa:2025pew}. For them, we find values of $d_{\text{reh}}$ up to $\mathcal{O}(10^4)$.
The total dilution factor is given by $d = d_\textrm{vac} \times d_\textrm{reh}$, thus even if the vacuum decay itself only results into a moderate dilution, its final effect can be sizeable for large values of $\alpha$.

These results are in line with those of the conformal dark sector model from~\cite{Goncalves:2025uwh}, for which the best fit point with respect to \nano data predicts a dilution $d_\text{reh}=1.15\cdot10^4$, and also with those of the conformal dark sector model from~\cite{Balan:2025uke}, for which the two best fit points presented predict a dilution $d_\text{reh}=400$ and $d_\text{reh}=9.86\times10^4$.
We report the dilution factors for the benchmark points analysed in~\cite{Costa:2025csj},~\cite{Costa:2025pew},~\cite{Goncalves:2025uwh} and~\cite{Balan:2025uke} in Table~\ref{tab:d_tot}\footnote{The values of $T_p$, $T_\text{reh}$ and $\alpha$ for the benchmark points contained in~\cite{Costa:2025csj} and~\cite{Costa:2025pew} differ slightly from those tabulated here because we are employing a more refined and numerically stabler computation of $P_f$, that will be incorporated in the public release of~\texttt{ELENA}.}.

\begin{table}[htb]
    \centering
    \begin{tabular}{|c|c|c|c|c|c|c|c|c|c|c|c|}
    \hline
    \multirow{2}{*}{} &
    \multicolumn{4}{c|}{\textit{Costa et al.}~\cite{Costa:2025csj}} & \multicolumn{4}{c|}{\textit{Costa et al.}~\cite{Costa:2025pew}} &
    \textit{Gon{\c{c}}alves et al.}~\cite{Goncalves:2025uwh} & \multicolumn{2}{c|}{\textit{Balan et al.}~\cite{Balan:2025uke}} \\ \cline{2-12}

    & \textbf{BP1} & \textbf{BP2} & \textbf{BP3} & \textbf{BP4} & \textbf{MAP} & \textbf{Bayes} & \textbf{MLE} & \textbf{Fast} &\textbf{Best fit point} & \textbf{Point A} & \textbf{Point B}\\
    \hline

    $T_p [\text{MeV}]$ & 9.60 & 18.1 & 51.2 & 38.7 & 0.74 & 0.68 & 0.11 & 14.9 & / & 2.28 & 1.21 \\ \hline
    $T_\text{reh} [\text{MeV}]$ & 53.8 & 108 & 312 & 120 & 4.46 & 10.6 & 4.05 & 39.7 & 11.7 & 16.8 & 55.9 \\ \hline
    $\alpha$ & 989 & $1.26\cdot10^3$ & $1.38\cdot10^3$ & 91.8 & $1.34\cdot10^3$ & $5.86\cdot 10^4$ & $2.05\cdot 10^6$ & 50.6 & $2.6\cdot 10^5$ & $4.7\cdot10^3$ & $9.48\cdot10^6$ \\ \hline
    $d_\text{reh}$ & 176 & 211 & 226 & 29.7 & 222 & $3.77\cdot10^3$ &$5.42\cdot10^4$ & 19.0 &$1.15\cdot10^4$ & 400 & $9.86\cdot10^4$ \\ \hline
    $d$ & 188 & 237 & 281 & 36.8 & 234 & $3.98\cdot10^3$ & $5.85\cdot10^4$ & 19.9 & / & / & / \\ \hline
    \end{tabular}
    \caption{Values of the percolation temperature $T_p$, the reheating temperature $T_\text{reh}$, the transition strength $\alpha$, the reheating contribution to the dilution factor $d_\text{reh}$ and the total dilution factor $d$ for the four BPs presented in~\cite{Costa:2025csj}, the five BPs presented in~\cite{Costa:2025pew} (the ``Slow'' BP is the same as BP1 in the previous study), the best fit point presented in~\cite{Goncalves:2025uwh} and for the two best fit points presented in~\cite{Balan:2025uke}.}
    \label{tab:d_tot}
\end{table}

In order to account for \nano data in the minimal dark sector model from~\cite{Costa:2025csj}, one thus generally expects a dilution of existing relic quantities of order $\mathcal{O}(10\ \text{--}\ 10^4)$.
In order to derive general results, and given the large range of (model dependent) values for the predicted dilution factors, we conduct a general analysis on the effect of  dilutions values $d$ on the parameter space of the leptogenesis model in Eq.~(\ref{eq:lagrangian}).

\begin{figure}[htbp]
    \centering
    \includegraphics[width=0.99\textwidth]{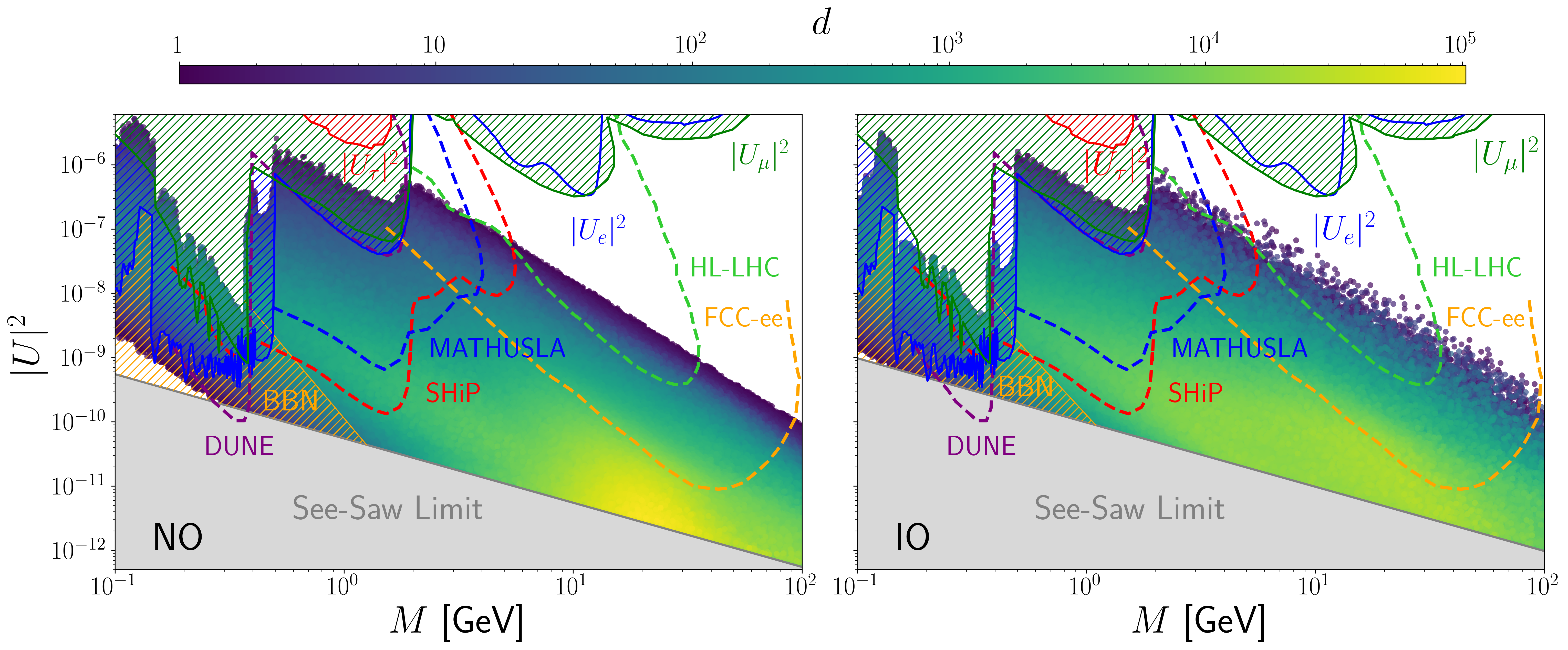}
    \caption{Leptogenesis parameter space that produces the observed BAU for different values of dilution $d$ in the model in Eq.~(\ref{eq:lagrangian}), for a normal (NO, left panel) and inverted (IO, right panel) ordering of light neutrino masses. Constraints from laboratory searches in the single flavour dominance limit are reported as blue ($|U_e|^2$), green ($|U_\mu|^2$) and red ($|U_\tau|^2$) dashed regions, and we exclude points that violate at least one of these. Below the See-Saw Limit, presented as solid region, the model cannot reproduce the light neutrino masses. The bound from BBN is given as a transparent grid, but no cut is performed on the points, to underline the fact that it is generally relaxed by the dilution of the HNL population, with the exact revised bounds depending on the specific dilution factor and reheating temperature values (see discussion in the main text). The expected sensitivities of DUNE, SHiP, MATHUSLA, HL-LHC and FCC-ee are shown as coloured dashed lines~\cite{Izaguirre:2015pga, Antusch:2016ejd, Antusch:2017pkq, Curtin:2018mvb, Pascoli:2018heg, SHiP:2018xqw, Ballett:2019bgd, Drewes:2019fou, Gorbunov:2020rjx, Drewes:2021nqr, Abdullahi:2022jlv}.}
    \label{fig:Max_cut}
\end{figure}

\titleprl{Effect of dilution on leptogenesis parameter space} We highlight how a testable leptogenesis model can provide insights on entropy injections in the early Universe in the $[1 \textrm{ MeV}, 130 \textrm{ GeV}]$ temperature range, and, vice-versa, how such new physics mechanisms constrain the parameter space for leptogenesis.
To study leptogenesis in the model in Eq.~(\ref{eq:lagrangian}), we perform a random sampling of the parameters adopting the priors previously described and, for any given dilution $d$, we accept values of the generated baryon asymmetry such that $|Y_B|\geq d\cdot Y_B^{\text{obs}}$ with $Y_B^{\text{obs}} \equiv n_b /s = 8.65 \times 10^{-11}$ the central value for the experimentally observed baryon yield~\cite{Planck:2018vyg}. We consider a lower bound on the required $Y_B$  since the value of the asymmetry can be generally reduced by modifying the CP-phases. Other than requiring agreement with neutrino oscillation data, we impose the current upper bounds on the active-sterile neutrino mixing in the single flavour dominance\footnote{The constraints on the active-sterile neutrino mixing coming from the experimental searches are on the individual mixings $|U_e|^2$, $|U_\mu|^2$ and $|U_\tau|^2$. We report in Fig.~\ref{fig:Max_cut} these individual constraints for reference, but highlight that the sum of mixings can lie above the individual flavour exclusions, as long as each flavour-specific constraint is complied with.}~\cite{Fernandez-Martinez:2023phj}. We also show an upper bound on the lifetime of each heavy neutral lepton state of 0.023 seconds~\cite{Boyarsky:2020dzc} to comply with BBN constraints. We point out that such bound will be generically weakened since the HNL population produced before the FOPT is diluted by the same mechanism as the BAU. Depending on the HNL mass and reheating temperature, some HNL can also be produced after the FOPT, for a discussion of HNL production in low reheating cosmologies see~\cite{Gelmini:2008fq}. Since the resulting effects depend on the model parameters, we report the standard BBN bound with the caveat that it would need to be reevaluated for each specific case, i.e. reheating temperature and dilution factor. We stress that such a BBN bound relaxation can possibly extend the parameter space testable by DUNE and SHiP. The results of our study are reported in Fig.~\ref{fig:Max_cut}, where we show the viable leptogenesis parameter space for different values of the dilution parameter $d$, for normal (NO, left panel) and inverted (IO, right panel) ordering of the light neutrino masses in the plane $|U|^2=\sum_\alpha|U_\alpha|^2$ vs $M$.
In general, we find that by increasing the dilution factor $d$ the parameter space restricts from all directions, such that the region of solutions becomes more and more localised in both masses and mixing values. Interestingly, the points that can produce the maximal lepton asymmetry are localised just above the seesaw line, for masses $M \simeq 20$ GeV ($M \simeq 30$ GeV) for the NO (IO). Thus, the restriction on the parameter space from below is less relevant, since it mostly affects a region where the model cannot account for neutrino data. On the contrary, the reduction of parameter space from all other directions is phenomenologically important, since it implies smaller mixings and a narrower range of masses to account for the BAU. It is worth noting that if a theory predicts a large value of $d$, this significantly narrows the parameter space and provides a prediction for where to experimentally search for heavy neutrinos.
The viable parameter space is reduced to a small region that shrinks to disappear around $d_\text{max}^\text{NO}=1.0\cdot10^{5}$ ($d_\text{max}^\text{IO}=3.0\cdot10^{4}$) for NO (IO), a value comparable with the dilutions expected in the models from~\cite{Costa:2025csj},~\cite{Goncalves:2025uwh} and~\cite{Balan:2025uke}.
Thus, we explicitly see that any FOPT taking place in the temperature range $[1 \textrm{ MeV}, 130 \textrm{ GeV}]$ with $\alpha_\text{NO} > 4.6 \times 10^6$ is incompatible with the low-scale leptogenesis mechanism with 2 RHN for a NO of active neutrino masses. For the IO case, the bound is stronger, with values of $\alpha_\text{IO} > 9.3 \times 10^5$ incompatible with this leptogenesis mechanism. For example, the maximum likelihood point (MLE) from~\cite{Costa:2025pew} and the second best fit point from~\cite{Balan:2025uke} predict values of dilution (see Table \ref{tab:d_tot}) that make the FOPT completely incompatible with the leptogenesis model in IO and NO, respectively. These bounds are actually conservative, since we are neglecting here the additional contribution from vacuum decay $d_\textrm{vac}$, that is more model dependent since it is related to the shape and temperature evolution of the finite-temperature scalar potential of the underlying theory.

\titleprl{Impact on experimental searches} The shrinking of parameter space observed in Fig.~\ref{fig:Max_cut} affects the testability of this model at planned experiments. In particular, we observe that the region of viable leptogenesis solutions is no longer testable by SHiP for values of $d_\textrm{NO}^\textrm{SHiP} > 3 \times 10^{3}$ ($d_\textrm{IO}^\textrm{SHiP} > 10^{4}$) for NO (IO) of light neutrino masses, corresponding in the context of FOPT to values of $\alpha > \alpha_\textrm{NO}^\textrm{SHiP} = 4.3 \times 10^4$ ($\alpha > \alpha_\textrm{IO}^\textrm{SHiP} = 2.2 \times 10^5$), and analogously for FCC with values $d_\textrm{NO}^\textrm{FCC} > 2 \times 10^4$ ($d_\textrm{IO}^\textrm{FCC} > 2.5 \times 10^4$), corresponding to $\alpha > \alpha_\textrm{NO}^\textrm{FCC} = 5.4 \times 10^5$ ($\alpha > \alpha_\textrm{IO}^\textrm{FCC} = 7.3 \times 10^5$). We again stress that these are conservative bounds since we are ignoring the further dilution term $d_\textrm{vac}$. Although it may seem counter-intuitive that the bound on future experiments in NO is more stringent than the one in IO, this depends on the fact that, in the sensitivity regions of SHIP and FCC, the parameter space reproducing the observed BAU in IO admits higher maximum dilution values than in NO (even if in the total parameter space $d_\text{max}^\text{NO}>d_\text{max}^\text{IO}$) due to the different $Y_B$ scaling with mass and mixing values. On the other hand, should HNL be observed in future experiments, the values of the physical masses and mixings in the parameter space of Fig.~\ref{fig:Max_cut} will determine the maximal amount of dilution that is compatible with the observed BAU; thus, assuming the Type-I seesaw with 2 RHN in Eq.~(\ref{eq:lagrangian}) as the only source of BAU will provide an upper bound on any entropy injection taking place in the temperature range $[1 \textrm{ MeV}, 130 \textrm{ GeV}]$, including (but not limited to) FOPT, whose transition strength would be bounded from above.

Notice that a strong experimental effort is also ongoing to search for low-scale dark sectors,~\cite{Antel:2023hkf,Alimena:2025rqb,Alimena:2025kjv} such as the ones that can be at the origin of the \nano signal. Thus, in case of a positive observation, the experimental feedback can also happen in the opposite direction, by singling out the amount of dilution expected from the new physics and correspondingly restricting the parameter space for successful leptogenesis.

\titleprl{Conclusion} We demonstrated that testable leptogenesis mechanisms can be used as powerful trackers of the early Universe thermal history in the wide range of temperatures $[1 \textrm{ MeV}, 130 \textrm{ GeV}]$. This study was motivated by the recent \nano evidence of a SGWB in the nanoHertz frequency range, but the underlying logic can be generalised to any new physics mechanism releasing entropy at some point in the considered temperature window. We adopted the minimal SM extended with 2 GeV-scale RHN as viable leptogenesis example model, even though the results can be generalised to further scenarios. In our case study, we demonstrated that there exists an upper bound on the strength of a FOPT, above which leptogenesis is incompatible with the observed BAU. Very significantly, such transition strength values are well within reach in some new physics models that can account for the \nano data, implying a strong impact on leptogenesis mechanisms from the low-scale new physics that is possibly at the origin of the nanoHertz SGWB signal. Even for FOPT models that predict more moderate values of the transition strength the impact on the leptogenesis parameter space can be important, significantly reducing the upper bound on the viable active-sterile mixings values, as well as the range of viable HNL masses.

On a longer timescale, assuming future experiments will be able to measure the parameters entering in a leptogenesis mechanism, the measured values will fix (or at least strongly restrict, depending on the uncertainties) the BAU value $Y_B$ generated during leptogenesis, opening a new observational window into the dynamics of the early Universe in the temperature range $[1 \textrm{ MeV}, 130 \textrm{ GeV}]$: if $Y_B > Y_B^{\text{obs}}$ one necessarily needs a dilution of the generated asymmetry before the BBN onset, which in turns requires new physics below the EW scale; if instead the generated BAU is compatible with the observed one, one can most likely assume a standard cosmological history and the validity of leptogenesis as the mechanism at the origin of the BAU.

\begin{acknowledgments}
\titleprl{Acknowledgments}
We thank Filippo Sala and Mauro Pieroni for valuable discussions and clarifications. L.~G. and M.~L. thank Fermilab for hosting them during a significant part of this project. The numerical results have been obtained by employing the Open Physics Hub HPC facility, hosted at the Department of Physics and Astronomy, University of Bologna.  L.~G.'s stay at Fermilab was founded by an International Mobility Scholarship for students of the ``Collegio Superiore'' of Alma Mater Studiorum - University of Bologna. L.G. acknowledges funding from the Deutsche Forschungsgemeinschaft (DFG): Project No. 386119226. M.~L. is funded by the European Union under the Horizon Europe's Marie Sklodowska-Curie project 101068791 — NuBridge. This work has been partly funded by the European Union under the Horizon Europe's  Project: 101201278 – DarkSHunt - ERC - 2024 ADG. Views and opinions expressed are however those of the author(s) only and do not necessarily reflect those of the European Union or the European Research Council Executive Agency. Neither the European Union nor the granting authority can be held responsible for them.
\end{acknowledgments}

\bibliographystyle{apsrev4-1}
\bibliography{bibliography}{}

\newpage
\onecolumngrid
\setlength{\parindent}{15pt}
\setlength{\parskip}{1em}
\newpage
\hypertarget{Supp_Mat}{}

\begin{center}
	\textbf{\large Boiling away asymmetries: low-scale phase transitions, gravitational waves and leptogenesis} \\
    \vspace{0.05in}
	{ \small \bf (Supplemental Material)}\\ 
	\vspace{0.05in}
    {Leonardo Grimaldi, Michele Lucente, Silvia Pascoli}
\end{center}

\section{Computation of the dilution factor due to vacuum decay}
To estimate the release of entropy due to the transition of the Universe from the false to the true vacuum state during a FOPT, we compute the vacuum entropy $S_\text{vac} = a^3 s_\text{vac}$, where $a$ is the scale factor and $s_\text{vac}$ is the entropy density defined as $s_\text{vac} = P_f s_f + (1-P_f) s_t$, where $s_f$ ($s_t$) is the entropy density of the false (true) vacuum and $P_f$ is the fraction of the Universe in the false vacuum. We are interested in the entropy change before and after the transition, thus we normalize the results to the entropy value at the transition critical temperature $T_c$:
\begin{equation}
    \Delta S (T, T_c) = \frac{S(T)}{S(T_c)} = \left(\frac{a(T)}{a(T_c)}\right)^3 \left(\frac{s(T)}{s(T_c)}\right)
    \label{Srel}.
\end{equation}

The first term in Eq.~\eqref{Srel} can be computed from the continuity equation
\begin{equation}\label{eq:continuity}
    \dot{\rho} = - 3 H \left[ \rho + p \right] \quad \Rightarrow \quad \left( \frac{a(T_1)}{a(T_2)} \right)^3 = \exp \left[ \int_{T_1}^{T_2} dT \frac{\frac{\partial \rho}{\partial T}}{\rho + p} \right],
\end{equation}
where $H$ is the Hubble parameter and we can estimate the total energy $\rho$ and pressure $p$ as
\begin{align}
    \rho(T) &= \rho_f(T) P_f(T) + \left(1 - P_f(T)\right) \rho_t(T),
    & \hspace{-5mm} \text{with} \hspace{1cm}
    \rho_{f,t}(T) &= V\!\left(\varphi_{f,t}(T),T\right)
    - \left. T \frac{\partial V}{\partial T}\right|_{\varphi = \varphi_{f,t}(T)},\\
    p(T) &= p_f(T) P_f(T) + \left(1 - P_f(T)\right) p_t(T),
    &  \hspace{-5mm} \text{with} \hspace{1cm}
    p_{f,t}(T) &= - V\!\left(\varphi_{f,t}(T),T\right),
\end{align}
where $V(\varphi, T)$ is the temperature-dependent effective potential of the full theory (SM plus new physics), $\varphi$ is the field driving the transition and the subscripts $f,t$ indicate the temperature-dependent values for the false and true vacua of $V$, respectively.

Notice that, given the entropy density is related to the thermal potential $V(\varphi,T)$ as
\begin{equation}
    s_{f,t}(T) = -\frac{\partial V(\varphi_{f,t}(T),T)}{\partial T},
\end{equation}
as long as there is a single phase ($P_{f,t} = 1$ and $d P_{f,t} / dT = 0$) the term in the integral simplifies to $(\partial s / \partial T) / s$ and the right-hand side of Eq.~(\ref{eq:continuity}) becomes $s(T_2) / s(T_1)$, resulting in
\begin{equation}
   \left. \Delta S (T_1, T_2) \right|_{\textrm{single phase}} = 1.
\end{equation}
However, this is generally not the case during the phase transition, where the system transitions from a vacuum state with larger ($\varphi_f$) to one with smaller ($\varphi_t$) entropy density, resulting in the entropy difference to be released into the thermal plasma.

\begin{figure}[htb]
    \centering
    \includegraphics[width=0.99\textwidth]{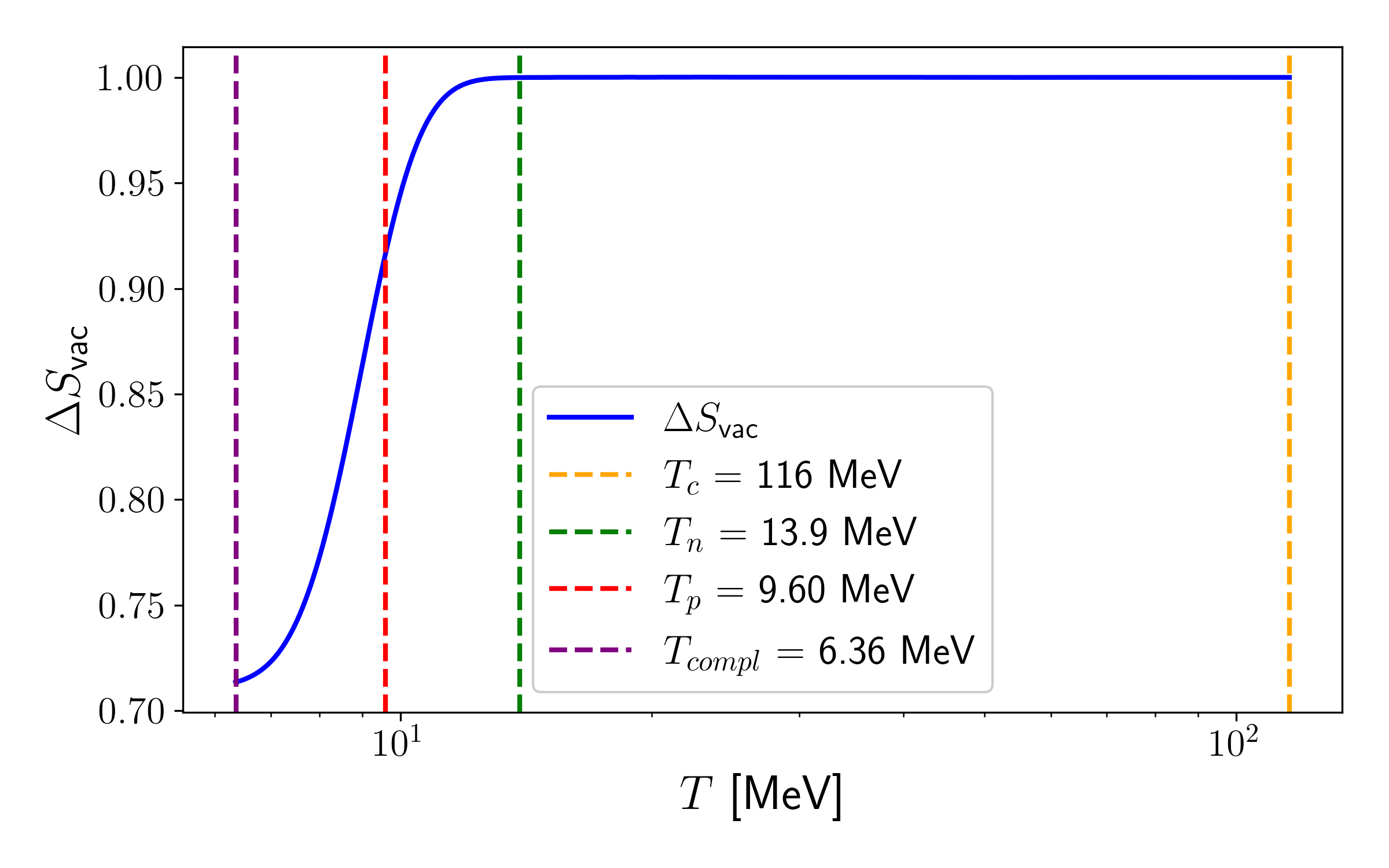}
    \caption{Plot of $\Delta S(T,T_c)$ as defined in Eq.~\eqref{Srel} for $T$ in the range between completion ($T_{\text{compl}}$) and critical ($T_{\text{c}}$) temperatures for the BP1 in~\cite{Costa:2025csj} (see also Table \ref{tab:d_tot}). For reference we signal also the nucleation ($T_n$) and percolation ($T_p$) temperatures.}
    \label{fig:S4}
\end{figure}

The second term in Eq.~(\ref{Srel}) can instead be computed as
\begin{equation}
    \frac{s(T)}{s(T_c)}=\frac{P_f(T)s_f(T)+[1-P_f(T)]s_t(T)}{s_f(T_c)},
\end{equation}
where the denominator results from the fact that $P_f = 1$ at the critical temperature.
We employ the public software \texttt{ELENA}~\cite{Costa:2025pew} to perform the above computations, and show as an example the entropy evolution for the benchmark point BP1 from~\cite{Costa:2025csj} in Fig.~\ref{fig:S4}. We can see that the total entropy is conserved down to the nucleation temperature $T_n$, when the nucleation of bubbles becomes significant, and decreases until the transition ends at the completion temperature $T_{\text{compl}}$.
Finally, the amount of dilution of the baryon yield due to this entropy release can be estimated as
\begin{equation}
    d_{\text{vac}}=1+\frac{S_\text{vac}(T_c)-S_\text{vac}(T_{\text{compl}})}{S_{\text{rad}}(T_c)},
    \label{eq:dcool}
\end{equation}
where $S_{\text{rad}}(T_c)$ is the radiation entropy at $T_c$.
It is convenient to rewrite \eqref{eq:dcool} as
\begin{equation}
    d_{\text{vac}}-1=\frac{S_\text{vac}(T_c)-S_\text{vac}(T_{\text{compl}})}{S_{\text{vac}}(T_c)}\frac{S_\text{vac}(T_c)}{S_{\text{rad}}(T_c)},
\end{equation}
such that the first term of the right-hand side is the fraction of entropy lost by the vacuum, $1-\Delta S_\text{vac}(T_\text{compl},T_c)$,
which for the case shown in Fig.~\ref{fig:S4} is 0.287.
The second term can be rewritten as $s_\text{vac}(T_c)/s_{\text{rad}}(T_c)$ since the scale factor is computed at the same temperature, where
\begin{equation}
    s_\text{rad}(T_c) = \frac{2\pi^2}{45} g_*(T_c) T_c^3,
\end{equation}
with $g_*$ the effective number of relativistic degrees of freedom. 
The results for all the BPs from~\cite{Costa:2025csj} and~\cite{Costa:2025pew} are reported in Table~\ref{tab:dBP}.

\begin{table}[htb]
    \centering
    \begin{tabular}{|c|c|c|c|c|c|c|c|c|c|}
    \hline
    \multirow{2}{*}{} &
    \multicolumn{4}{c|}{\textit{Costa et al.}~\cite{Costa:2025csj}} & \multicolumn{5}{c|}{\textit{Costa et al.}~\cite{Costa:2025pew}} \\ \cline{2-10}

    & \textbf{BP1} & \textbf{BP2} & \textbf{BP3} & \textbf{BP4} & \textbf{MAP} & \textbf{Bayes} & \textbf{MLE} & \textbf{Slow} & \textbf{Fast} \\
    \hline

    $d_{\text{vac}}$ & 1.064 & 1.125 & 1.241 & 1.242 & 1.055 & 1.055 & 1.078 & 1.064 & 1.050 \\ \hline
    \end{tabular}
    \caption{Value of the dilution factor due to vacuum decay for the BPs contained in~\cite{Costa:2025csj} and~\cite{Costa:2025pew} (the ``Slow'' BP is the same as BP1).}
    \label{tab:dBP}
\end{table}

\end{document}